\documentclass[aps,apl,psfig,amstex,showpacs,twocolumn]{revtex4-1}

\def\ind#1{{_{\mathrm{#1}}}}

\usepackage{amssymb}
\usepackage{amsmath}
\usepackage{graphicx,color}

\begin{document}

\title{Magnetoelectric coupling in a ferroelectric/ferromagnetic chain revealed by ferromagnetic resonance}

\author{A. Sukhov$^{1,2}$, P.P. Horley$^{2}$, C.-L. Jia$^{1,3}$, J. Berakdar$^{1}$}

\affiliation{
  $^{1}$ Institut f\"ur Physik, Martin-Luther Universit\"at Halle-Wittenberg, 06120 Halle (Saale), Germany\\
  $^{2}$ Centro de Investigaci\'{o}n en Materiales Avanzados (CIMAV S.C.), Chihuahua/Monterrey, 31109 Chihuahua, Mexico\\
  $^{3}$ Key Laboratory for Magnetism and Magnetic Materials of the Ministry of Education, Lanzhou University, Lanzhou 730000, China}


\begin{abstract} Understanding the multiferroic coupling  is one of the key issues in the field  of  multiferroics.
As shown here theoretically, the ferromagnetic resonance (FMR) renders possible an access to the magnetoelectric coupling coefficient
in composite multiferroics. This we evidence by a  detailed analysis and numerical calculations of FMR in an unstrained chain
of  BaTiO$_3$ in the tetragonal phase in contact with Fe, including the effect of depolarizing field.
 The spectra of the absorbed power in FMR are found to be sensitive to the orientation of the interface electric polarization and to an applied static electric field. Here we propose a method for measuring the  magnetoelectric coupling coefficient by means of FMR.
\end{abstract}

\maketitle

\textit{Introduction.-}
Materials with multi  ferroic (magnetic, electric, and/or elastic) orders, called multiferroics (MF), have attracted increased attention again  \cite{SpFi05,Fi05,EeMa06,RaSp07}, mainly due to the discovery that the notoriously small multiferroic  coupling in bulk matter
 may well be increased by a controlled engineering
of low dimensional compounds,   opening  thus  the way for the design of qualitatively new device concepts
 \cite{BiBa08,VaHo10,Zavaliche-11}. For instance,
magnetoelectricity allows for the control of
 magnetism with an electric field which has a large potential for
 environmentally friendly sensorics  and   spintronics applications with low-energy consumption.
 The progress in this field depends critically on the understanding  of the magnetoelectric (ME)
coupling and on developing methods to assess its properties. This is particularly so, as  currently several  ME coupling mechanisms   are
 being discussed, e.g. in Refs. [\onlinecite{WaLi09,VaHo10,LaSr11,LuKi11}].
On the other hand, an established approach for probing the ferromagnetic response is the ferromagnetic resonance \cite{Gr46,Za46,Vo66,Fa98} in which the sample is usually subjected  to crossed static and  time-dependent magnetic fields.
  %
   Hence, it is natural and timely to envisage a possible mapping of the
  multiferroic dynamics in an FMR setup with the aim to draw conclusions on the nature of the ME coupling and relate it to the  FMR signal.
  In fact, FMR  has been experimentally shown   \cite{goennenwein_11,goennenwein_10} to be sensitive to acoustic waves in ferromagnetic-ferroelectric structure.
   To our knowledge, multiferroic FMR, as suggested below has not yet been realized experimentally for the chosen interface, though several studies are known for multiferroic interfaces with other types of ME-coupling \cite{BiKo01,LiLi11}.
  To  be specific, we focus on a special class of ME coupled materials,  so-called composite MFs \cite{VaHo10,NaBi08}, that may  be synthesized
  from a wide range of materials that, when composed together yield a strong ME coupling  and stable ferroelectric (FE) and ferromagnetic (FM) orders  at room temperatures.
  An example that has been studied intensively, theoretically and experimentally is
  BaTiO$_3$(BTO)/Fe \cite{DuJa06,Binek-07,Tsymbal-08,Taniyama-09,SuJi10,Bibes-11,MeKl11,Taniyama-12,CaJu09,bob12,EPL12}.
The ME coupling in this system is predicted to be an interfacial effect and relatively high \cite{DuJa06,CaJu09,FeMa10,LeSa10}.
 An experimental evidence is presented in [\onlinecite{MeKl11}].\\
For a reliable prediction, a realistic modelling is indispensable since the ME coupling is relatively weak  in total (due to its interfacial nature)
and the MF dynamics is governed by a series of interrelated effects, as shown below.
The route followed here is based on a combination of the Landau-Lifshits-Gilbert and the Ginzburg-Landau dynamics using the total MF free energy $F\ind{\Sigma}$  density \cite{SuJi10,JiSu11,HoSu12}, including the  ferroelectric  $F\ind{FE}$, the ferromagnetic $F\ind{FM}$, and the part
 $E\ind{INT}$  that involves the ME interface coupling.
 $F\ind{FE}$  includes the energy densities corresponding to the Ginzburg-Landau-Devonshire potential $F\ind{GLD}$\cite{Gi49,De49}, to the spatial inhomogeneity  of the order parameter $F\ind{GE}$ \cite{Hl06}, the depolarizing field contribution $F\ind{DF}$\cite{RaAh07}, the dipole-dipole interaction $F^{\mathrm{FE}}_{\mathrm{DDI}}$, as well as the applied external electric field $F\ind{AEF}$.
 $F\ind{FM}$ \cite{Coey10} incorporates  the nearest-neighbor exchange interaction $F\ind{EXC}$, the anisotropy contribution $F\ind{ANI}$, the FM dipole-dipole interaction $F^{\mathrm{FM}}_{\mathrm{DDI}}$ and the energy density of the interaction with an external magnetic field $F\ind{AMF}$.
 Realistic parameters corresponding to
bulk BaTiO$_3$ in the tetragonal phase and bulk Fe are employed.
Our simulation show that the FMR spectra of the absorbed power are indeed sensitive to the FE order as controlled by
  an applied static electric field, evidencing thus the access of FMR to the ME coupling.
 The key finding of this study is a practical proposal to  estimate   the  magnetoelectric coupling coefficient from the FMR spectra.
    With the refinement in the spatiotemporal
  resolution of the FMR technique we expect the multiferroic FMR to play a vital role in uncovering the details of ME coupling.\\
\textit{Theoretical model.-}
Our treatment is based on the Ginzburg-Landau phenomenology, i.e. we study the dynamics of coarse-grained order parameters
 ($\vec{P}_i$ and $\vec{M}_j$) of the FE/FM chain, that  result from an averaging of  the microscopic quantities over an appropriate cell. These cells form in our calculations
 the sites $i$ or $j$ for the local  $\vec{P}_i$ or  $\vec{M}_j$.
At the interface  (site $i=j=1$) of  the FE/FM composite  the mobile spin-polarized electrons in the FM rearrange as to screen  the electric polarization
  at the FE part \cite{Zh99}, leading thus in effect to a local ME coupling of
  the interface  magnetization $\vec{M}_1$ with the interface polarization  $\vec{P}_1$ that can
   be expressed \cite{CaJu09} as $E\ind{INT}=\lambda \vec{P}_1\cdot \vec{M}_1$, where $\lambda$ is the ME coupling coefficient.
The  FE polarization vector $\vec{P}_i$ develops in time  according to the Landau-Khalatnikov (LKh) equation\cite{LaKh54}
\begin{equation}
\displaystyle \gamma_{\nu}\frac{d \vec{P}_i}{dt} = -\frac{\delta F\ind{\Sigma}}{\delta \vec{P}_i},
\label{eq_1}
\end{equation}
with the FE relaxation constant $\gamma\ind{\nu}=2.5\cdot 10^{-5}$~[Vms/C]\cite{MaHl08}.
The   magnetization dynamics $\vec{M}_j$ is governed by the Landau-Lifshitz-Gilbert (LLG) equation\cite{LaLi35,Gi55}
\begin{eqnarray}
\displaystyle \frac{d\vec{M}_j}{dt}&=&-\frac{\gamma}{1+\alpha_{\mathrm{FM}}^2}\left[ \vec{M}_j \times \vec{H}^{\mathrm{LEF}}_j(t) \right]\\ \nonumber
&&-\frac{\alpha\ind{FM}\gamma }{(1+\alpha_{\mathrm{FM}}^2)M\ind{S}} \left[ \vec{M}_j \times \left[ \vec{M}_j \times \vec{H}^{\mathrm{LEF}}_j(t) \right] \right],
\label{eq_2}
\end{eqnarray}
where $\gamma$ is the gyromagnetic ratio and $\alpha\ind{FM}$ is a Gilbert damping constant. The local effective field is defined as $\vec{H}^{\mathrm{LEF}}_j(t)\equiv -\frac{\delta F\ind{\Sigma}}{\delta \vec{M}_j}$.
The FMR power absorbed by the chain we infer from (cf. e.g., \cite{Us06,SuUs08})
\begin{equation}
\displaystyle P\ind{FMR}=- \mu_0 a^3_{\mathrm{FM}} \sum_j\frac{1}{N_{\mathrm{T}}T}\int_0^{N_{\mathrm{T}}T} \vec{M}_j(t)\cdot \frac{\partial \vec{H}_{\Sigma}}{\partial t} dt,
\label{eq_3}
\end{equation}
where $\mu_0$ is the vacuum magnetic constant, $a\ind{FM}$ is the  cell size, $N\ind{T}$ and $T=2\pi/\omega$ are the number and the period of the external magnetic field cycles, respectively. The sum runs over magnetization sites $j$. The total magnetic field applied to the system is $\vec{H}\ind{\Sigma}(t)=H\vec{e}\ind{z}+H_0\cos\omega t \vec{e}\ind{y}$, whereas $H_0\ll H$.\\
It the following the imaginary part of the transverse magnetic susceptibility $\chi''$ will be calculated, for which $\chi'' \sim \frac{P\ind{FMR}}{\mu_0 M\ind{S}H_0\omega}$ holds\cite{Vo66,Fa98}.\\
\textit{Contact of ultrathin FE and FM ($E=0$).-}
\begin{figure}
\includegraphics[scale=.9]{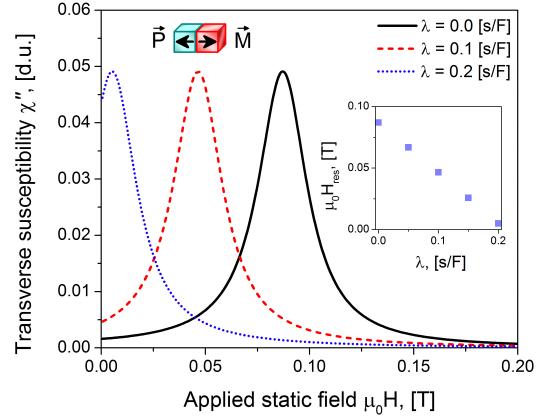}\label{fig_1}
\caption{FMR spectra for the multiferroic chain of $N\ind{FE}=1$ and $N\ind{FM}=1$-sites. BaTiO$_3$ has the tetragonal phase with the following coefficients in the free energy density (eq. (\ref{eq_4})) $\alpha\ind{FE}=-2.77\cdot 10^7$~[Vm/C]\cite{Hl06}, $\beta\ind{FE}=1.7\cdot 10^8$~[Vm$^5$/C$^3$]\cite{Hl06} and $P\ind{S}=0.265$~[C/m$^2$]\cite{Hl06}. The parameters of bulk Fe assumed in the calculations (eq. (\ref{eq_5})) for this figure are: $K_1=4.8\cdot 10^4$~[J/m$^3$]\cite{Coey10}, $M\ind{S}=1.71\cdot 10^6$~[A/m]\cite{Coey10} and $\alpha\ind{FM}=0.1$. For the time-dependent magnetic field it is taken $\mu_0H_0=28\cdot 10^{-3}$~[T], $\omega/(2\pi)=4\cdot 10^9$~[Hz]. Inset shows the dependence of resonance peaks on $\lambda$.}
\end{figure}
First, we numerically model a contact of a single FE site and a single FM site in zero E-field.
 Employing the FE free energy density for the tetragonal phase of BaTiO$_3$\cite{SuJi10}
\begin{equation}
\displaystyle F^{\mathrm{N=1}}_{\mathrm{FE}}=\frac{\alpha\ind{FE}}{2}P^2_{\mathrm{z}}+\frac{\beta\ind{FE}}{4}P^4_{\mathrm{z}}
\label{eq_4}
\end{equation}
and  the FM free energy density for a uniaxial crystal in an external magnetic field \cite{SuJi10}
\begin{equation}
\displaystyle F^{\mathrm{N=1}}_{\mathrm{FM}}=-\frac{K_1}{M^2_{\mathrm{S}}}\left(M\ind{z}\right)^2-\mu_0\vec{M}\cdot \vec{H}\ind{\Sigma}(t).
\label{eq_5}
\end{equation}
we calculated the spectra shown in Fig. 1.
 The position of the peak for the case of zero ME coupling follows  from\cite{Ki48}
$\frac{1}{\gamma}\omega|_{\mathrm{res}}(\lambda=0)=\mu_0 H^{\mathrm{LEF}}|_{\mathrm{res}}=\frac{2K_1}{M^2_{\mathrm{S}}}M\ind{z}+\mu_0 H\ind{res}$
according to which we find $\mu_0 H\ind{res}(\lambda=0)\approx 0.087$~[T] (Fig. 1) \cite{foot1}.
For a finite ME coupling constant the resonance field is modified by the anisotropy induced by the screened polarization
\begin{equation}
\displaystyle \mu_0 H\ind{res}=\frac{1}{\gamma}\omega|_{\mathrm{res}}(\lambda\neq 0)-\frac{2K_1}{M^2_{\mathrm{S}}}M\ind{z}+\lambda P\ind{z}.
\label{eq_7}
\end{equation}
This relation reveals that the resonance condition depends on the orientation of the surface polarization.
 In the simulations the spectra are calculated at equilibrium with the effect of opposite orientation of the polarization and magnetization. For a  finite $\lambda$ the magnetization is aligned along $z$ and the polarization $P\ind{z}$ is negative. This results in a shift of the resonance peaks towards smaller fields (Fig. 1, $\lambda \neq 0$). The shift of resonance fields remains linear as a function of $\lambda$ (inset of Fig. 1).\\
We also note, that the ME coupling does not affect the intensity and the width of the spectra, similarly to the effect of the uniaxial anisotropy axis oriented along $z$ axis in macro-spin FM nanoparticles\cite{SuUs08}.\\
\textit{Contact of ultrathin FE and FM ($E\neq 0$).-}
Considering the MF chain in the presence of an electric field $E$ which acts  directly on the electrically active part of the chain,  we expect for $E\parallel z$ that the resonances shift to weaker magnetic fields (eq. (\ref{eq_7})). To favor the stability of the magnetization along the $z$ direction,
  we increase the frequency of the oscillating magnetic field which lowers    the intensity of $\chi''$.
Similarly to  Fig. 1, a finite $\lambda$ leads to a shift of the resonance field towards smaller fields ($\lambda=0$ is not shown in Fig. 2),
this is also predicted by a very recent study~\cite{bob12}.
The nonzero electric field applied parallel to the MF chain acts on the FE polarization and indirectly changes the ME coupling energy enlarging the resonance fields (eq. (\ref{eq_7}) for the field $\vec{E}=E\vec{e}\ind{z}$).\\
Let us  define the shift of the resonance position from eq. (\ref{eq_7}) for a finite electric field with a reference to zero E-field
\begin{eqnarray}
\displaystyle \mu_0\Delta H &\equiv& H\ind{res}(E\neq 0)-H\ind{res}(E=0)= \nonumber \\
&& \lambda \Delta P\ind{z} \equiv \lambda \left[P\ind{z}(E\neq 0) - P\ind{z}(E=0)\right],
\label{eq_8}
\end{eqnarray}
and plot $\mu_0\Delta H$ against $\Delta P\ind{z}$ (Fig.3, left), we can determine from the slope of the obtained dependence the ME coupling parameter $\lambda=0.202$~[s/F], the original value of which was set to $0.2$~[s/F].
The analysis of the polarization states shows that the abrupt increase of the resonance field for $E=5\cdot 10^6$~[V/m] (Fig. 2) refers to the reversal of the polarization from the antiparallel to the parallel arrangement with respect to the magnetization orientation in the FM part. Since re-polarization of the FE  modifies significantly the system and creates an energy jump at the interface responsible for a parallel orientation of FE and FM sites, it is natural to expect that the corresponding resonance point will deviate from the linear dependence, which is proved by Fig. 3 (left panel). Therefore, a linear fit of the spectral data should be performed  for the points obtained for the electric fields that do not result in re-polarization of the ferroelectrics ($E<5\cdot 10^5$~[V/m]). If the spectral points measured for high fields  were included into the fitting data set, the resulting values of $\lambda$ can exceed the actual value considerably.
Notwithstanding the difficulties of FE polarization measurements \cite{SaTo30} during the FMR experiment,
the proposed procedure gives a transparent method for obtaining the magnitude of the ME-coupling coefficient \cite{foot2}.
\begin{figure}
\includegraphics[scale=0.8]{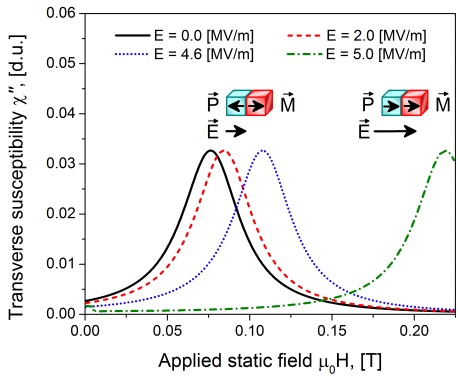}
\label{fig_2}
\caption{FMR spectra for the multiferroic chain of $N\ind{FE}=1$ and $N\ind{FM}=1$-sites. All parameters are adopted from Fig. 1, except the ME coupling parameter $\lambda=0.2$~[s/F] and the frequency of the applied magnetic field $\omega/(2\pi)=6\cdot 10^9$~[Hz].}
\end{figure}

\begin{figure}
\includegraphics[scale=0.8]{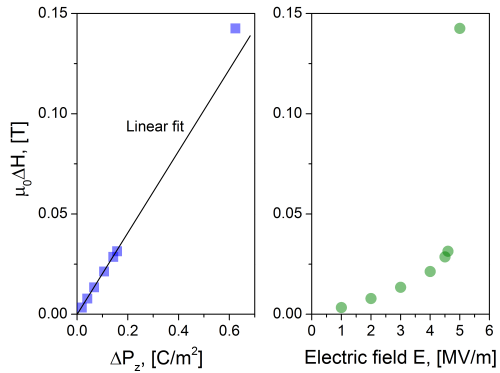}
\label{fig_3}
\caption{Shifts of resonance positions calculated according to expression (\ref{eq_8}) for different applied electric fields. Values of FE polarizations are calculated for the resonance magnetic fields presented in Fig. 2. Resonance curves for the fields $E=\{1.0; 3.0; 4.0; 4.5\}\cdot 10^6$~[V/m] are not shown in Fig. 2.}
\end{figure}

\textit{Thin FE/FM contact for $E\neq 0$.-}
For a thicker MF system the total FE energy density additionally includes interactions of the neighboring sites\cite{SuJi10} and the dipole-dipole interactions $F^{\mathrm{FE}}_{\mathrm{DDI}}$\cite{HoSu12} $ F^{\mathrm{N>1}}_{\mathrm{FE}}=\sum_i \Big(\frac{\alpha\ind{FE}}{2}P^2_{\mathrm{z} i}+\frac{\beta\ind{FE}}{4}P^4_{\mathrm{z} i} +\frac{\kappa\ind{FE}}{2}(P_{\mathrm{z} i+1}-P_{\mathrm{z} i})^2-P_{\mathrm{z} i}E\ind{z}\Big)+F^{\mathrm{FE}}_{\mathrm{DDI}}$. The FM energy density (eq. (\ref{eq_5})) is supplemented by the exchange interaction and the FM dipole-dipole interaction $F^{\mathrm{FM}}_{\mathrm{DDI}}$\cite{HoSu12} $ F^{\mathrm{N>1}}_{\mathrm{FM}}=\sum_j \Big(-\frac{A}{a^2_{\mathrm{FM}}M^2_{\mathrm{S}}}\vec{M}_j\cdot\vec{M}_{j+1}-\frac{K_1}{M^2_{\mathrm{S}}}\left(M_{\mathrm{z} j}\right)^2-\mu_0\vec{M}_j\cdot \vec{H}\ind{\Sigma}(t)\Big)+F^{\mathrm{FM}}_{\mathrm{DDI}}$. The ME coupling  in our case  is limited to the vicinity of the FE/FM interface. Hence, we  expect  the influence of ME coupling on MF dynamics to be less pronounced  as compared with the single-state case (Fig. 2), which is  evidenced by Fig. 4.
Nevertheless, the spectral lines are clearly distinguished for FE switching from antiparallel to parallel orientation regarding the direction of magnetization vectors in FM part. The FE switching occurs for lower field values in comparison with single-state case because the presence of interaction between the sites lowers the barrier of the Ginzburg-Landau-Devonshire potential. Applying the procedure outlined above, we plotted  (inset to Fig. 4) the variation of peak position versus the averaged ferroelectric polarization. Remarkably, the obtained
points fits well to a straight line for the case of low polarization values corresponding to a non-switched ferroelectric layer.
 The points related to the switched value diverge from a linear scaling and were not considered. The slope of the linear fit obtained from the plot is 0.01376, which is lower than the value of $\lambda = 0.2$ [s/F]. This effect most probably appears because the magnetoelectric coupling is limited to a single interface and then propagates through
a chain of five sites at each side of the device, resulting in a smaller variation of FMR spectra.
A straightforward step consisting in a multiplication of the fitted value of $\lambda$ by the number of sites yields the value of 0.0688 [s/F] that is three times smaller than $\lambda$. We think that it is possible to find a proper re-normalization constant from geometric considerations, which, however, go beyond the scope of this letter. The most important result is that the peak position /
polarization variation plots present the same linear dependence as that observed for the single-site system, which endorse  the proposed method for the measurement of the magnetoelectric coupling for thicker composite multiferroic systems.
\begin{figure}
\includegraphics[scale=1.]{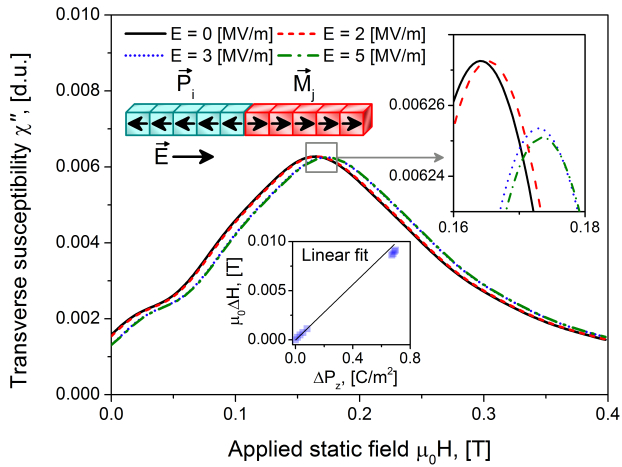}
\label{fig_4}
\caption{FMR spectra for a  chain of $N\ind{FE}=5$ and $N\ind{FM}=5$ sites,  cell size: $a\ind{FE}=a\ind{FM}=5\cdot 10^{-9}$~[m].
The applied field frequency is  $\omega/(2\pi)=30\cdot 10^9$~[Hz], FM exchange stiffness is $A=2.1\cdot 10^{-11}$~[J/m] \cite{Coey10}, and $\lambda=0.2$~[s/F].  Other parameters  are as in Fig. 1.
 The FE coupling strength is calculated as $\kappa\ind{FE}=G_{11}/a^2_{\mathrm{FE}}$, where $G_{11}=51.\cdot 10^{-11}$~[Jm$^3$/C$^2$]\cite{Hl06}.
 Insets show the peak on an enlarged scale resolving
  the states with non-switched and switched ferroelectric layer. The linear fit of the resonance shifts involves the non-switched polarization.}
\end{figure}

\textit{Remarks and conclusions.-}
As shown above,  the FMR spectra of the absorbed power of an unstrained thin composite mutliferroic chain  depend critically on the
 ME coupling. The peaks of resonance absorption are sensitive to the magnitude and the orientation of the FE polarization vector (Fig. 1) in the absence of an electric field. Applying a static electric field changes the value of $P\ind{z}$  and causes a shift in  the peak position according to  (eq. (\ref{eq_7})). Figs. 2 and 3 demonstrate how in our case the ME coupling can be accessed. At first, the static electric field is applied along the direction of the FE minima such that it  shifts sizably the peaks relative to the field-free case.
The shifts of the fields $\mu_0 \Delta H$ (eq. \ref{eq_8}) are plotted against the measured FE polarization, yielding the ME coupling coefficient as the slope of the $\mu_0 \Delta H (\Delta P\ind{z})$ dependence (Fig. 3, left panel).
The plot of $\mu_0 \Delta H(E)$ (Fig. 3, right panel) is nonlinear, hence, the ME coupling can only be determined for known $P\ind{z}(E)$ dependence. We note,  the $\mu_0 \Delta H(E)$ dependence will be different for the rhombohedral phase\cite{HoSu12} of the FE \cite{SuHo_unpubl}.

Based on the results in (Figs. 3, 4),  on the estimates made for the ME coupling coefficient \cite{SuJi10,HoSu12}, and  on the calculations for the elongated MF chains \cite{JiSu11} we suggest to choose a MF contact consisting of a thin FM layer (electrode) and a thick FE. The thin FM layer serves to avoid additional peaks in the FMR spectra due to spin waves as well as to avoid a decay of the ME coupling in long FMs, whereas the thick FE part stabilizes the FE polarization for measuring $P\ind{z}$ (Fig.3, left).
Note, for highlighting  the ME dynamics dominated by the interfacial screening charges, the principal axis of Fe is rotated by 45$^{\circ}$ with respected to BTO[100], we have then a good lattice match ($\sim 1.4 \%$) for the epitaxial growth of bcc Fe on the tetragonal BTO.
\acknowledgments{
The authors gratefully acknowledge the support of the German Research Foundation by the grant SU 690/1-1 and SFB 762, the grant of CONACYT as Basic Science Project 129269 (Mexico), the National Natural Science Foundation of China (Grant No. 11104123) {and the National Basic Research Program of China (Grant No. 2012CB933101).}}


\begin{thebibliography}{30}
\bibitem{foot1}{The magnetization is supposed to be aligned by a resonant magnetic field along the $z$ axis ($M\ind{z}\approx M\ind{S}$)}.
\bibitem{foot2}
A straightforward method in an experiment would be to plot $\mu_0 \Delta H$ as a function of the applied electric field $E$. The resonance shifts against the electric field are shown in Fig. 3 (right), which has a nonlinear character and can be explained due to the solution for the FE polarization from $\frac{\delta F\ind{\Sigma}}{\delta P\ind{z}}=0$, yielding $P^3_{\mathrm{z}}+\frac{\alpha\ind{FE}}{\beta\ind{FE}}P\ind{z}+\frac{1}{\beta\ind{FE}}\left(\lambda M\ind{S}-E\right)=0.$ This eq. gives a strongly nonlinear solution for $P\ind{z}(E)$, which, assuming the relation (\ref{eq_8}), suggests a nontrivial dependence of $\mu_0 \Delta H(E)$ even for the case of low fields, when the FE polarization is not reversed yet.
\end{thebibliography}
\end{document}